\UseRawInputEncoding
\documentclass[12pt, twoside, epsf]{article}

\usepackage{color,graphicx,times}
\usepackage{amsmath, amssymb, graphics}

\newcommand{\mathsym}[1]{{}}
\newcommand{\unicode}[1]{{}}

\begin{document}

\date{}

\title{\bf A Construction for Clifford Algebras}

\author{Louis H. Kauffman \\
  Department of Mathematics, Statistics and Computer Science \\
  University of Illinois at Chicago \\
  851 South Morgan Street\\
  Chicago, IL, 60607-7045}

\maketitle
  
\thispagestyle{empty}

\noindent {\bf Abstract.} {\it This paper is dedicated to the memory of Zbigniew Oziewicz, to his
generosity, intelligence and intensity in the search that is science and mathematics.}The paper explains a basic construction producing Clifford algebras inductively, starting with a base algebra $A$ that is associative and has an involution.  The basic construction always produces associative algebras and can be iterated indefinitely. The basic construction is generalized to a group theoretic construction where a group $G$ acts on the algebra $A.$ This group theoretic construction generates Clifford algebras and matrix algebras. The paper then studies applications of this algebra to mathematical and physical examples.\\



\section{Introduction}
This paper begins with a basic construction that produces Clifford algebras inductively, starting with a base algebra $A$ that is associative and has an involution. This basic construction is an analog of the
Cayley-Dickson Construction that produces the complex numbers, quaternions and octonions starting from the real numbers. Our basic construction always produces associative algebras and can be iterated
an indefinite number of times. We generalize the basic construction to a group theoretic construction where a group $G$ acts on the algebra $A,$ and show how this group theoretic construction is related to matrix algebras. The paper then concentrates on applications of this algebra to mathematical and physical examples.\\
 
In Section 2 we give an analog of the Cayley-Dickson Process (that generates quaternions, octonions and a hierarchy of non-associative algebras) for the construction of Clifford algebras. This basic construction
starts with an associative algebra $A$ with involution ($a \longrightarrow a^{\star}$) and produces a new associative algebra $\hat{A}$ with involution that contains a new element $\eta$ such that 
$\eta^2 = 1$ and for all $a$ in $A,$ $\eta a \eta = a^{\star}.$ Iteration of the basic construction produces Clifford algebras. Since this interaction always produces associative algebras, this basic construction can be
regarded as a foundation for the construction of Clifford algebras. Section 2 gives the basic construction and discusses examples. Section 2.2 shows how the basic construction can be seen to unfold logically from 
a way to formalize the square root of minus one as a discrete oscillation. Section 2.3 gives a group theoretic generalization of the basic construction where the involution is replaced by the action of an arbitrary group
on the initial algebra $A.$ Section 2.3 reviews the Cayley-Dickson Construction.\\

We mention here a key special case of the Basic Construction, discussed in Section 2. Let $A = R \times R$ denote all ordered pairs of real numbers $[a,b]$ with the involution $[a,b]^{\star} = [b,a].$
Pairs are added and multiplied coordinatewise. $\hat{A}$ contains $\eta$ as above with $\eta [a,b] \eta = [a,b]^{\star} = [b,a].$ Thus if $e=[-1,1],$ then we have $\eta e \eta = [1,-1] = - e,$ whence
$e^2 = \eta^2 = 1$ and $e \eta = - \eta e.$ Here is a small Clifford algebra and we see that if $i = e \eta,$ then $i^2 = e\eta e\eta = e (-e) = - e^2 = -1.$ Thus $i = e \eta$ is a square root of minus one.
We can regard $e \eta$ as a formalization of the idea that the square root of minus one corresponds to an oscillation between minus one and plus one. This is discussed in the body of the paper.
In the case of $A = R \times R,$ we have that $\hat{A}$ consists in the linear combinations $[a,b] + [c,d]\eta,$ and in Sections 2 and 3 we show that this is isomorphic with the ring of $2 \times 2$ matrices over $R.$\\

It is important to point out that the involution in the basic construction is a representation of the cyclic group of order two. This means that we assume that 
$(ab)^{\star} = a^{\star} b^{\star}.$ The order of multiplication is preserved under a product. The conjugation fundamental to the Cayley-Dickson Construction reverses the order of multiplication. This makes the main difference between these two constructions.\\

Section 3 is an introduction to a process algebra of iterants seen as applications of the group theoretic construction of Section 2.3.  This section shows how iterants give an alternative way to do $n \times n$ matrix algebra and how the ring of all $n \times n$ matrices can be seen as a faithful representation of an iterant algebra based on the cyclic group of order $n.$  Section 3 ends with a number of classical examples including representations for quaternion algebra.  \\

Section 4 discusses the structure of the Dirac equation and how the nilpotent and the Majorana operators arise naturally in this context. This section provides a link between our work and the work on nilpotent structures and the Dirac equation of Peter Rowlands \cite{Rowlands}. We end this section with an expression in split quaternions for the the Majorana Dirac equation in one dimension of time and three dimensions of space. The Majorana Dirac equation can be written as follows:
$$(\partial/\partial t + \hat{\eta} \eta \partial/\partial x + \epsilon \partial/\partial y + \hat{\epsilon} \eta \partial/\partial z - \hat{\epsilon} \hat{\eta} \eta m) \psi = 0$$ where $\eta$ and $\epsilon$ are the simplest generators of iterant algebra with $\eta^{2} = \epsilon^{2} = 1$ and $\eta \epsilon + \epsilon \eta = 0,$
and $\hat{\epsilon}, \hat{\eta}$ form a copy of this algebra that commutes with it. This combination of the simplest Clifford algebra with itself is the underlying structure of Majorana Fermions, forming indeed the underlying structure of all Fermions.  In Section 4 we apply the nilpotents method to the Majorana Dirac Equation and give actual real solutions to the equation. These solutions inevitably make direct use of the
Majorana Fermion Clifford algebra. This shows that it is not just a formal relationship that Fermions and Majorana Fermions are related by the algebraic transformation between Fermion and Clifford algebra. We end the paper with a specific special case in one dimension of space and one dimension of time and its relationship with the Feynman Checkerboard \cite{Feynman,KN:Dirac}.\\

\section{The Basic Construction}
In this paper we explain an inductive construction of Clifford algebras that is an analog to the well-known Cayley-Dickson construction that produces the complex numbers from the real numbers, the quaternions from 
the complex numbers and the octonions from the quaternions in a uniform inductive sequences. Our construction for Clifford algebras always produces an associative algebra from a given associative algebra, and so provides a new proof that standard Clifford algebras are associative. Other insights are available from this construction as we shall discuss in the course of the paper. In this section,  we give the definition of the 
construction and outline its main properties.\\

\subsection{The construction of $\hat{A}$ from $A.$}
For our purposes, a Clifford algebra $\cal{A}$ is an associative algebra that is a module over a given associative algebra $A$ generated by linearly independent elements $\{ \eta_1, \eta_2, \cdots, \eta_n \}$
such that $$\eta_{k}^{2} = 1$$ for each $k=1,\cdots n$ and
$$\eta_{i} \eta_{j} + \eta_{j} \eta_{i} = 0$$ whenever $i \ne j.$ That such algebras exist is well-known, but we shall give a proof that constructs them inductively and that generalizes to a class of algebras that 
we call {\it iterant algebras}. The first construction that we give shows the existence of Clifford algebras. In particular, we show that they are associative algebras. The construction we use  is in direct analogy with the Cayley-Dickson construction \cite{Dickson,Chatelin,Conway,Dray,Baez} that gives rise to the quaternions and the octonions and to other algebras as well. In the course of our constructive proof, we show that every Clifford algebra has an involution $a \longrightarrow a^{\star}$ so that $(a^{\star})^{\star} = a$ and $(ab)^{\star} = a^{\star} b^{\star}$ for every $a$ and $b$ in the Clifford algebra. In particular, we have that $1^{\star} = 1$ and that $\eta_{i}^{\star} = - \eta_{i}$ for each special element $\eta_{i}.$\\

Let $A$ be an associative algebra with unit $1$ and assume that A is endowed with an involution $a \longrightarrow a^{\star}$ satisfying the following properties.
\begin{enumerate}
\item $1^{\star} = 1.$
\item $(a^{\star})^{\star} = a$ for all $a$ in $A.$
\item $(ab)^{\star} = a^{\star} b^{\star}$ for all $a$ and $b$ in $A.$
\end{enumerate}
Note that in the third property, the application of the involution to a product is the product of the involutions in the same order of multiplication.
We say that $A$ is an {\it associative algebra with involution.}\\

Given such an algebra, we define a new algebra $\hat{A}$ with involution,  by adjoining an independent element $\eta$ with the following properties.
\begin{enumerate}
\item $\hat{A}$ consists in the set of all $a + b \eta$ with $a$ and $b$ in $A$ so that $a + b \eta = c + d \eta$ if and only if $a=c$ and $b = d$ in $A.$
\item $(a + b\eta) + (c + d \eta) = (a+c) + (b+d) \eta$ for all $a,b,c,d$ in $A.$
\item $\eta^{\star} = - \eta$ and $(a + b \eta)^{\star} = a^{\star} - b^{\star} \eta.$
\item $\eta^{2} = 1.$
\item $\eta a = a^{\star} \eta$ for all $a$ in $A.$ Hence $\eta a \eta = a^{\star}$ for all $a$ in $A.$
\item The specific rule for multiplication of elements of $\hat{A}$ is given by the formula
$$(a+ b \eta)(c + d \eta) = (ac + bd^{\star}) + (ad + bc^{\star}) \eta.$$  
\end{enumerate}

We now prove that $\hat{A}$ is itself an associative algebra with involution, showing that the construction can be iterated.\\

\noindent {\bf Lemma.} $\hat{A}$ is associative.\\

\noindent {\bf Proof.} $$((a + b \eta)(c + d \eta))(f + g\eta) = ( (ac + bd^{\star}) + (ad + bc^{\star}) \eta ) ( f + g \eta)$$
$$= ( (ac + bd^{\star})f + (ad + bc^{\star})g^{\star} ) + ((ac + bd^{\star})g +  (ad + bc^{\star})f^{\star})\eta$$
$$= ( acf + bd^{\star}f + adg^{\star} + bc^{\star}g^{\star} ) + (acg + bd^{\star}g +  adf^{\star} + bc^{\star}f^{\star})\eta$$

$$(a + b \eta)((c + d \eta)(f + g\eta)) =(a + b\eta) ((cf + d g^{\star}) + (cg + d f^{\star})\eta)$$
$$= (a(cf + d g^{\star}) + b (cg + d f^{\star})^{\star})  + ( a (cg + d f^{\star}) + b  (cf + d g^{\star})^{\star} )\eta $$
$$= (acf + ad g^{\star} + b (c^{\star} g^{\star} + d^{\star} f)  + ( a c g + a d f^{\star} + b c^{\star} f^{\star} + b d^{\star} g )\eta $$

Thus $$((a + b \eta)(c + d \eta))(f + g\eta) = (a + b \eta)((c + d \eta)(f + g\eta))$$ for any $a,b,c,d,f,g$ in $A.$ 
\noindent This completes the proof. 
\fbox{}\\

\noindent {\bf Lemma.} $\hat{A}$ is an associative algebra with involution.

\noindent {\bf Proof.} We have defined the involution on  $\hat{A}$ by the formula $(a + b \eta)^{\star} = a^{\star} - b^{\star} \eta.$
We need to verify that for elements $z$ and $w$ in $\hat{A},$  $(zw)^{\star} = z^{\star} w^{\star}.$  Accordingly, let $z = a + b \eta$ and $w = c + d \eta .$
Then 
$$z^{\star} w^{\star} = (a^{\star} -  b^{\star} \eta)(c^{\star} -  d^{\star} \eta)$$
$$=(a^{\star} c^{\star} + b^{\star} (d^{\star})^{\star}) + (-a^{\star}d^{\star}  - b^{\star}(c^{\star})^{\star})\eta$$
$$=(a^{\star} c^{\star} + b^{\star} d) - (a^{\star}d^{\star}  + b^{\star}c) \eta,$$
and
$$(zw)^{\star} = ((ac + bd^{\star}) + (ad + bc^{\star}) \eta)^{\star}$$
$$= (ac + bd^{\star})^{\star} - (ad + bc^{\star})^{\star} \eta$$
$$=(a^{\star} c^{\star} + b^{\star} d) - (a^{\star}d^{\star}  + b^{\star}c) \eta.$$
This completes the proof. \fbox{}\\

These two lemmas imply that the Clifford construction of $\hat{A}$ from $A$ can be iterated. Let $A^{[n]}$ denote the iteration of the construction $n$ times so that
$\hat{\hat{A}} = A^{[2]}$ and $\hat{\hat{\hat{A}}} = A^{[3]}.$  We see that in $A^{[n]}$ we have adjoined elements $\eta_1, \eta_2, \cdots \eta_n$ so that
\begin{enumerate}
\item $\eta_k ^{2} = 1$ for each $k,$  and $\eta_{k} ^{\star} = - \eta_k .$
\item $\eta_i \eta_j \eta_i = \eta_j ^{\star} = - \eta_j$ whenever $i > j.$ Hence $$\eta_i \eta_j + \eta_j \eta_i = 0$$ whenever $i \ne j, $ and indeed
$$\eta_i \eta_j \eta_i = \eta_j ^{\star} = - \eta_j$$ whenever $i \ne j.$
\end{enumerate}
We have all the products of the form $\eta_{\vec{i}} = \eta_{i_1}\eta_{i_2}\eta_{i_3} \cdots \eta_{i_{k}}$  where $\vec{i} = (i_1, i_2, i_3, \cdots , i_k)$ where these indices denote
a subset of $\{ 1,2,3,\cdots, n \}$ with $k$ distinct elements. It is sufficient to assume that $i_1 < i_2 < i_3 < \cdots < i_n.$ The elements of $A^{[n]}$ are linear combinations of 
products of the for form $\eta_{\vec{i}}$ with coefficients in $A.$ Products follow from associativity and the relations and commutation relations for the generators $\eta_i.$ These are
Clifford Algebras over the base algebra $A.$ \\

We can examine how a given $A^{[n]}$ works by examining the new elements in it. Thus we have the following examples.
\begin{enumerate}
\item Let $R$ denote the real numbers with the identity involution so that $r^{\star} = r$ for all real numbers $r.$ Then $\hat{R} = \{ t + x \eta \}$ with 
$\eta^2 = 1$ and $(t + x \eta)^{\star} = t - x \eta,$ with $t$ and $x$ real numbers. Thus we have, for $z = t + x \eta,$ $$z z^{\star} = t^2 - x^2.$$ The reader will 
recognize this as the space-time metric for one dimenision of space and one dimension of time. $\hat{R}$ can be identified as the Minkowski Plane.
\item With $R$ as above, we can form $\hat{\hat{R}} = R^{[2]}.$ This algebra is generated by $1, \eta, \lambda, \eta \lambda$ over $R$ where $\eta^2 = 1, \lambda^2 =1$ and
$\eta \lambda = - \lambda \eta.$ From this we see that if $\gamma = \eta \lambda$ then $\gamma^2 = \eta \lambda \eta \lambda = \eta \eta^{\star} = - \eta^2 = -1.$
This algebra, with two elements of square one and another of square minus one, and each pair anti-commuting, is often called the {\it split quaternions}. It can be converted into the quaternions by tensoring it
with a copy of the complex numbers $C$ that commutes with these special elements. Then $i \eta$ and $i \lambda$ have square minus one, and we can identify
$$I = i \eta, J = i \lambda, K = \lambda \eta$$ and find that $$I^2 = J^2 = K^2 = IJK = -1$$ so that the resulting algebra $R^{[2]} \otimes C$ is isomorphic with the quaternions. 
\item If we begin with the complex numbers $C=\{ a + bi | a,b \in R, i^2 = -1 \}$ and form $\hat{C}$ we add one element $\eta$ and take the involution on $C$ to be conjugation so that 
$r^{\star}= r $ when $r$ is real and $i^{\star} = - i.$ Then in $\hat{C}$ we have $\eta i \eta = i^{\star} = - i,$ whence $\eta i = - i \eta.$ Making this extension from the complex numbers has $i$ anti-commuting
with the special element $\eta.$ On the other hand it is easy to see that $(i \eta)^2 = 1$ and that $i \eta$ anti-commutes with $i$ and $eta.$ Thus we have that $\hat{C}$ is isomorphic to
$R^{[2]}.$ The initial idea that the square root of minus one is an entity that commutes with the other numbers no longer holds in these extended systems.
\item In $R^{[3]}$ there are three new special elements. Call them simply $\eta_1, \eta_2, \eta_3$ as we have indicated above. Then we can see another appearance of the quaternions in the form 
$I = \eta_ 2 \eta_1, J = \eta_3 \eta_2, K = \eta_1 \eta_3.$ The reader will have no difficulty in verifying that $I^2 = J^2 = K^2 = IJK = -1.$ Thus the quaternions are a subalgebra of this much larger algebra of
dimension eight over the real numbers. 

\item Let $A= R \times R = \{ [a,b] | a,b \in R\}$ with operations $$[a,b][c,d] = [ac, bd], [a,b] + [c,d] = [a+c, b+d], [a,b]^{\star} = [b,a].$$ In $\hat{A}$ we adjoin an element $\eta$ such that $\eta^2 = 1$ and
$\eta[a,b]\eta = [a,b]^{\star} = [b,a].$ Thus $\eta [a,b] = [b,a]\eta.$ $\hat{A}$ consists in the set of $[a,b] + [c,d]\eta$ with these operations. It is not hard to see that  $\hat{A}$ is isomorphic with 
$M_{2}(R),$ the ring of $2 \times 2$ matrices over $R.$ The isomorphism is given by 
$$[a,b] + [c,d]\eta \longrightarrow 	\left(\begin{array}{cc}
			a&c\\
			d&b
			\end{array}\right).$$
Note that in $\hat{A} = \hat{R \times R}$ we have a copy of the Clifford algebra $\hat{\hat{R}}$ using the elements $e = [-1,1]$ and $\eta,$ since $e^2 = \eta^2 = 1$ and $e\eta = - \eta e.$
The element $e \eta$ has square equal to minus one, and corresponds to the matrix
$$\left(\begin{array}{cc}
			0&-1\\
			1&0
			\end{array}\right).$$
At this point it is natural to ask where in this construction do we get the action of $i$ on pairs of real numbers in the form $i(a,b) = (-b,a).$ One answer is that this comes from $i(a + ib) = -b + ia.$
Another is in the form of the product of a matrix times a column vector:
$$\left(\begin{array}{cc}
			a&c\\
			d&b
			\end{array}\right)
			\left(\begin{array}{c}
			x\\
			y
			\end{array}\right) = 
			\left(\begin{array}{c}
			ax+cy\\
			dx+by
			\end{array}\right)$$
			Note that the end result corresponds to $[ax+cy, dx+by] = [a,b][x,y] + [c,d][y,x].$
We define $$(A + B\eta).C = AC + BC^{\star}$$ as the analogue of the product of a matrix $M = A + B\eta$ times a vector $C$,
Then $$(e \eta).[a,b] = [-1,1][b,a] = [-b, a],$$ and we retrieve the usual action of $i$ on points in the Cartesian plane.
That we can construct the full $2 \times 2$ matrix algebra using our construction, suggests that a generalization of the construction could construct arbitrary matrix algebras. This is the case and will be discussed below and  in 
Section 3  of this paper.\\

\item{\bf Clifford algebra and Fermion algebra.}  In two by two matrix algebra, we can take
$$ \epsilon = \left(\begin{array}{cc}
			-1&0\\
			 0&1
			\end{array}\right),
			 \eta = \left(\begin{array}{cc}
			0&1\\
			1&0
			\end{array}\right) = a + b.$$
 
Here 
$$a = \left(\begin{array}{cc}
			0&0\\
			1&0
			\end{array}\right), b = \left(\begin{array}{cc}
			0&1\\
			0&0
			\end{array}\right).$$
Thus 
$$ab = \left(\begin{array}{cc}
			0&0\\
			0& 1
			\end{array}\right), ba = \left(\begin{array}{cc}
			1&0\\
			0&0
			\end{array}\right)$$
so that $$a^2 = b^2 = 0,$$
$$a + b = \eta,$$
$$a - b = \epsilon \eta,$$
$$ab + ba = 1,$$
$$ab - ba = \epsilon.$$

More abstractly, suppose that we have a Clifford algebra generated by elements $\epsilon$ and $\eta$ with $\epsilon^2 = \eta^2 = 1$ and $\epsilon \eta + \eta \epsilon = 0.$
Then we can define new elements $a$ and $b$ by the equations
$$\eta = a + b,$$
$$\epsilon \eta = a - b.$$
This means that 
$$a = \frac{1}{2}(1 + \epsilon) \eta,$$
$$b = \frac{1}{2}(1 - \epsilon) \eta,$$
from which it follows that 
$$a^2 = b^2 = 0, ab + ba = 1.$$
Note that we are given that the starting Clifford algebra is associative and so further identities such as 
$$aba = a, bab = b, abab = ab, baba = ba$$
follow easily from the given identities.
We call an associative algebra generated by $a,b$ with 
$$a^2 = b^2 = 0, ab + ba = 1$$
a {\it Fermion algebra} since the annihilation, creation algebra for Fermions in quantum theory satisfies these identities.
We see here that Clifford algebras (with an even number of generators)  and Fermion algebras are interchangeable via the above transformations.
This fact has been used by writers on Clifford algebras, \cite{S} since it is useful to have projector properties 
such as $(ab)(ab) = ab.$\\

\noindent {\bf Lemma.} The involution on the algebra generated by $\{ a,b \}$ with
$$\eta = a + b,$$ 
$$\epsilon \eta = a - b$$
 is given by the formulas
$a^{\star} = b,$
$b^{\star} = a.$

\noindent {\bf Proof.} 
Note that $\eta^{\star} = \eta$ and $(\epsilon \eta)^{\star} = - \epsilon \eta.$
Thus 
$$a^{\star} + b^{\star} = a + b,$$
$$a^{\star} - b^{\star} = -a + b.$$ From this it follows that $a^{\star} = b$ and $b^{\star} = a.$
This completes the proof.\\

Note that if $A$ is any associative algebra with involution, then the two-fold application of our construction, $\hat{\hat{A}},$ is an extension of $A$ by the algebra generated by $\eta$ and $\epsilon$ as above.
Thus we can regard $\hat{\hat{A}}$ as an extension of $A$ by the Fermion algebra as we have described it above. This relationship of Fermion algebra and Clifford algebra will
be used later in this paper and will also be the subject of a sequel to this paper.\\

\noindent{\bf Remark.} The above construction of Fermion algebra from Clifford algebra occurs without invoking an extra commuting square root of negative unity. It is common in physical applications
to use a parallel consttruction involving $i$ where $i^2 = -1$ and $i$ commutes with all elements of the algebra. One can then define $\psi = \frac{1}{2}(\eta + i \epsilon)$ and 
$\psi^{\dagger} = \frac{1}{2}(\eta - i \epsilon).$ It follows that $\psi^{2} = (\psi^{\dagger})^{2} = 0$ and $\psi \psi^{\dagger} + \psi^{\dagger} \psi = 1,$ and one has a Fermion algebra with complex conjugation constructed in relation to a Clifford algebra. See Sections 4 and 5 for a discussion of these constructions in relation to Majorana Fermions.

\end{enumerate}

\subsection{A Precursor to the Clifford Construction}
\begin{figure}
     \begin{center}
     \begin{tabular}{c}
     \includegraphics[width=6cm]{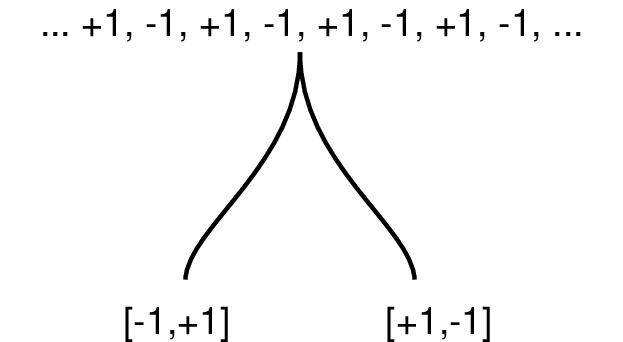}
     \end{tabular}
     \caption{\bf A Basic Oscillation}
     \label{Figure 1}
\end{center}
\end{figure}

 The idea for the construction $\hat{A}$ comes from an attempt to formalize the idea that $i$ can be regarded as an oscillation between plus and minus one.  In this view, $i$ is the fixed point of the recursion
 $T(x) = -1/x$ that oscillates for real values such as $+1$ or $-1.$ See Figure~\ref{Figure 1}.
 
 If we regard $i = \{ \pm 1\}$ so that in multiplying 
$i \times i$ there is a time shift so that $\{ \pm 1\}$ becomes $\{ \mp 1\},$ then upon multiplying we would have $$i \times i = \{ \pm \}\{\pm 1\}= \pm \mp 1  = -1$$
Algebraically, we can say that $$\{ x \} \{ y \} = x y^{\star}$$ where $\star$ is an appropriate involution so that $$\pm 1^{\star} =  \mp 1.$$ In the case where we identify  $\pm 1 = [+1, -1]$ and $ \mp 1 = [-1, +1],$ the natural involution is 
$[a,b]^{\star} = [b,a]$. In our Clifford construction we have extended the algebra so that the involution is represented by a formula of the type $ x^{\star} = \eta x \eta.$ We can now point out that in the constraint of 
producing an associative algebra, this is an inevitable feature.\\

Suppose we start with an associative algebra A with involution and wish to produce a new associative algebra with new elements $\{ x \}$ where $x$ is in $A$ and $\{ x \}\{ y \}=x y^{\star}.$ The next lemma shows
what further multiplication rules are needed to obtain a new associative algebra consisting of elements $x + \{ y\},$\\

\noindent {\bf Lemma.} Letting A be an algebra with involution, extend $A$ as indicated above with the following rules for multiplication. Then the extended algebra is associative.
\begin{enumerate}
\item $\{ x \}\{ y \}=x y^{\star}.$
\item $x \{ y \}=\{ x y \}.$
\item $\{x\} y = \{ x y^{\star} \}.$
\end{enumerate}

\noindent {\bf Proof.} We will check that $(\{ x \}\{ y \})\{ z \} = \{ x \}(\{ y \}\{ z \})$ and leave the rest of the proof to the reader.
$$(\{ x \}\{ y \})\{ z \} = xy^{\star}\{ z \}  =\{ xy^{\star} z \}.$$
$$\{ x \}(\{ y \}\{ z \}) = \{ x \}y z^{\star} = \{ x (y z^{\star})^{\star} \} = \{ x y^{\star} z \}.$$
This completes the proof. \fbox{}\\

Now let $\eta = \{1\}.$ Then for any $x$ in $A$ we have $$x \eta = x\{1\} = \{x\},$$ $$\eta^{2} = \{1\}\{1\} = 1.$$ Thus
$$\eta x \eta = \eta \{x\} = \{ 1\} \{x\} = x^{\star}.$$
We have shown that this formalization of the square root of negative unity as an oscillation is ismorphic with the Clifford Construction.\\

\subsection{The Group Theoretic Construction}
In the Clifford Construction, we had an involution acting on the initial algebra $A.$ This is the same as saying that the cyclic group of order two acts on $A.$ We now generalize to the case of an
arbitrary group $G$ acting on the algebra $A.$ We will give specific examples in this section and later in the paper as well.\\

We shall have an algebra $A$ acted upon by a group $G$ with the  action denoted by $a^{g}$ for $a$ in the algebra and $g$ in the group.
This generalizes our previous notation $a^{\star}$ for action of the involution on the element $a.$\\

To say that $A$ is acted upon by $G$ means that $(a^{h})^{g} = a^{gh}$ for all $a$ in $A$ and $g,h$ in $G.$ Note that this is formally the same as the functional equation
$g(h(a)) = (gh)(a).$\\

We extend to an algebra that includes the elements of the group itself via adding relations $g a = a^{g} g$  for $a \in A$ and $g \in G.$ Call the new algebra $A^{[G]}.$
Thus we have the following rules for multiplication:
\begin{enumerate}
\item It is given that $G$ and $A$ are associative and that for $g,h \in G$ and $a,b \in A,$  words formed using both algebras are associative. Thus $(ag)b = a(gb)$ and similarly for other mixed combinations.
Furthermore, we take by definition the formula $ga = a^{g} g.$
\item $(a^{h})^{g} = a^{gh}.$ Note that this corresponds to the identity $(gh)a = g(ha)$ since $$(gh)a = a^{gh} (gh)$$ and $$g(ha) = g(a^{h} h) = (ga^{h})h = ((a^{h})^{g} g)h = (a^{h})^{g} (gh).$$

\item Note how the associativity of the group theoretic construction works:
$$ (a g) (( b h) ( c k)) = (ag) ((bc^{h}) hk) = a ((bc^{h})^g g(hk)) = a b^{g} (c^{h})^{g} g(hk)$$
while
$$ ((a g) ( b h)) (c k)) = (a b^{g} gh)(ck) = a b^{g} c^{gh} (gh)k$$
and since
$$c^{gh} = (c^{h})^{g},$$ associativity follows from the associativity of $A$ and the associativity of the group multiplication.
\end{enumerate}

\noindent There is no need to iterate this construction, but if we take the group
to be an $n$-fold  product of groups of order two, then  $A^{[G]} = A^{[n]},$ the $n$-fold iterated Clifford Construction.\\

There are many examples for this group construction. Here we will discuss the {\it iterant} construction by which I mean $(R^{n})^{[G]}$ where $R^{n}$ denotes $n$-tuples of real numbers 
$[a_1,a_2, \cdots, a_n ]$ with coordinate-wise multiplication and addition. For the iterant construction it is given that there is a representation of $G$ to $S_n ,$ the permutation group on $n$-letters
and that $G$ acts through this representation on the $n$-tuples in $R^{n},$ permuting them accordingly. Thus in our previous example, we had $[a,b]^{\star} = [b,a]$ and this is a representation of the cyclic group of order two that interchanges the coordinates of the vector. Note that if $G$ is a finite group, then an element in $(R^{n})^{[G]}$ can be written as
$$\sum_{g \in G} v_{g} g$$ where $v_{g} \in R^n$ and $g \in G$ and 
$$ v_{g} gv_{h} h =  v_{gh} gh$$ where $v_{gh} = v_{g} (v_{h})^{g},$
 In this sense $(R^{n})^{[G]}$ is a generalization of the group ring $R[G]$ of $G$ whose elements are of the form $\sum_{g \in G} a_{g} g$ where
$a_g \in R.$ In fact, the group ring $R[G]$ is isomorphic to $(R^{1})^{[G]}$ where $G$ acts trivially on $R.$\\ 

For $M = \sum_{g \in G} v_{g} g$ and $w \in R^{n},$  define $$M. w = \sum_{g \in G} v_{g} w^{g}.$$ This is the analog of the action of a matrix $M$ on a column vector $w.$

In Sections 3 and 4 we will discuss examples of the group theoretic and iterant constructions. Much more is possible and will be explored in subsequent work.\\

\subsection{The Cayley-Dickson Construction}
Since, by this Clifford construction, we have encountered the quaternions in a number of ways as the construction is performed, it is natural to ask whether there is also an encounter with the 
Octonions \cite{Baez,Dray,Conway}. As is well known, the Octonions, or Cayley numbers, are not associative and so cannot appear directly in constructions of the kind we have outlined here, since we always make associative 
algebras. Nevertheless, the Octonions do appear from the Cayley-Dickson Construction \cite{CD}, and this is a good place to recall how that construction is done.\\

$$(a + Jb)(c + Jd) = (ac - d \bar{b}) + J (cb + \bar{a} d)$$
$$\overline{a + Jb} = \bar{a} - J b$$

The key point about the Cayley-Dickson Construction in contrast to our Clifford Construction is that the conjugation on the algebra is an {\it anti - homomorphism} of the algebra so that 
$\overline{XY} = \overline{X} \overline {Y}$ for all algebra elements $X$ and $Y.$ On this account, one finds that applying the Cayley-Dickson construction to the quaternions or more generally to a non-commutative algebra will lead to a non-associative algebra. In this way, the Octonions are not associative and other algebras that emerge from interating the construction are also non-associative. Nevertheless, there is 
a strong analogy between the Cayley-Dickson Construction and our construction for Clifford algebras. This analogy should be further pursued.\\

 \section{Iterants, Discrete Processes and Matrix Algebra}
 Recall from Section 2.3 that we have defined the {\it iterant} construction as follows: Iterants consist in $(R^{n})^{[G]}$ where $R^{n}$ denotes $n$-tuples of real numbers 
$[a_1,a_2, \cdots, a_n ]$ with coordinate-wise multiplication and addition. For the iterant construction it is given that there is a representation of $G$ to $S_n ,$ the permutation group on $n$-letters
and that $G$ acts through this representation on the $n$-tuples in $R^{n},$ permuting them accordingly. Thus we had $[a,b]^{\star} = [b,a]$ and this is a representation of the cyclic group of order two that interchanges the coordinates of the vector. Note that if $G$ is a finite group, then an element in $(R^{n})^{[G]}$ can be written as
$$\sum_{g \in G} v_{g} g$$ where $v_{g} \in R^n$ and $g \in G$ and 
$$ v_{g} gv_{h} h =  v_{gh} gh$$ where $v_{gh} = v_{g} (v_{h})^{g}.$\\ 

We begin in this section by discussing the iterant construction for  the algebra $$\mathbb{V}ect_{n}(\mathbb{R})= (R^{n})^{[C_{n}]}$$ where $C_{n}$ denotes the cyclic group of order $n.$ We consider $n=2$ and $n=3$ specifically, and we indicate the isomorphism
 $(R^{n})^{[C_{n}]} = M_{n}(R)$ of these iterant rings with the rings of $n \times n$ matrices over $R.$\\
 
The primitive idea behind an iterant is a periodic time series or 
waveform. The elements of the waveform
can be any mathematically or empirically well-defined objects. We can regard
the ordered pairs $[a,b]$ and $[b,a]$ as abbreviations for the waveform or as two points
of view about the waveform ($a$ first or $b$ first). Call $[a,b]$ an {\em iterant}. 
One has the collection of
transformations of the form $T[a,b] = [ka, k^{-1}b]$ leaving the product $ab$ invariant.
This tiny model contains the seeds of special relativity, and the iterants contain the seeds of 
general matrix algebra.
For related discussion see \cite{SS,SRF,SRCD,IML,KL,BL,Para,LOF}.
\bigbreak 

 Define products and sums of iterants as follows
 $[a,b][c,d] = [ac,bd]$  and $[a,b] + [c,d] = [a+c,b+d].$
 The operation of juxtapostion of waveforms is multiplication
 while $+$ denotes ordinary addition of ordered pairs. These operations are natural 
with respect to the 
 structural juxtaposition of iterants:
 $$...abababababab...$$
 $$...cdcdcdcdcdcd...$$ Structures combine at the points where they correspond.  
Waveforms combine at the times where they correspond. Iterants combine in juxtaposition.
\bigbreak

 If $\bullet$ denotes any form of binary compositon for the ingredients 
($a$,$b$,...) of
 iterants, then we can extend $\bullet$ to the iterants themselves by the 
definition
 $[a,b]\bullet[c,d] = [a\bullet c,b\bullet d]$.

 The appearance of a square root of minus
 one unfolds naturally from iterant considerations.  The shift operator 
$\eta$ on iterants is defined by the equation $\eta[a,b] = [b,a]\eta$ with $\eta^2 = 1.$  This is obtained forming the Clifford Construction of Section 2 to 
$R \times R.$
Sometimes it is 
convenient to think of
 $\eta$ as a delay opeator, since it shifts the waveform $...ababab...$  
by one internal
 time step. Now define  $$i = [-1,1]\eta $$
 We see at once that $ii = [-1,1]\eta [-1,1]\eta = [-1,1] [1,-1] \eta^2 = [-1,1] [1,-1] = [-1,-1] = -1.$ 
Thus  $ii=-1.$    Here we have 
described $i$  as the combination of the waveform 
 $\epsilon = [-1,1]$  and the temporal shift operator $\eta.$
By writing $i = \epsilon \eta$ we recognize an active version of the waveform that shifts temporally when it is observed.
\bigbreak

Now we show how all of matrix algebra can be formulated in terms of iterants.
\bigbreak
 
 \subsection{Matrix Algebra Via Iterants}
Consider a waveform of period three.
$$\cdots abcabcabcabcabcabc \cdots$$
Here we see three natural iterant views (depending upon whether one starts at $a$, $b$ or $c$).
$$[a,b,c],\,\,\, [b,c,a], \,\,\, [c,a,b].$$ The appropriate shift operator is given by the formula
$$[x,y,z]S = S[z,x,y].$$ Thus, with $T = S^{2},$
$$[x,y,z]T = T[y,z,x]$$ and $S^{3} = 1.$
With this we obtain a closed algebra of iterants whose general element is of the form $$[a,b,c] + [d,e,f]S + [g,h,k]S^{2}$$ where $a,b,c,d,e,f,g,h,k$ are real or complex numbers.
This algebra is $\mathbb{V}ect_{3}(R) = (R^{3})^{[C_{3}]}$ where the scalars are in the real numbers $R.$ Let $M_{3}(R)$ denote the $3 \times 3$ matrix algebra over $R.$ We have the 
\smallbreak 

\noindent {\bf Lemma.} The iterant algebra  $\mathbb{V}ect_{3}(R)$ is isomorphic to the full
$3 \times 3$ matrix algebra $M_{3}(R).$
\smallbreak

\noindent {\bf Proof.} See \cite{Iterants}. \fbox{}\\

\noindent {\bf Remark.} By the clear generalization of this example to cyclic groups of arbitrary finite order, it is easy to see that the algebra $(R^{n})^{[C_{n}]}$ is isomorphic to the algebra of 
$n \times n$ matrices over $R$, $M_{n}(R).$ Other groups can be treated similarly, and we can represent the group theoretic iterant construction for any finite group as a full algebra of matrices 
$M_{n}(R),$ where $n$ is the order of the group $G.$ See \cite{Iterants}. \\

\section{The Dirac Equation and Majorana Fermions}
 We now construct the Dirac equation. The algebra underlying this equation has the same properties as the  creation and  annihilation algebra for fermions. It is by way of this algebra that we will come to the Dirac equation. If the speed of light is equal to $1$ (by convention), then energy $E$, momentum $p$ and mass $m$ are
related by the (Einstein) equation $$E^2 = p^2 + m^2.$$ Dirac constructed his equation by looking
for an algebraic square root of $ p^2 + m^2$ so that he could have a linear operator for $E$ that would take the same role as the Hamiltonian in the Schr\"{o}dinger  equation. We will get to this operator by first taking the case where $p$ is a scalar (we use one dimension of space and one dimension of time.).
Let $E = \alpha p + \beta m$ where $\alpha$ and $\beta$ are elements of a a possibly non-commutative,
associative algebra. Then $$E^2 = \alpha^2 p^2 + \beta^2 m^2 + pm(\alpha \beta + \beta \alpha).$$
Hence we will satisfy $E^2 = p^2 +m^2$ if $\alpha^2 = \beta^2 = 1$ and
 $\alpha \beta + \beta \alpha = 0.$ This is our familiar Clifford algebra pattern and we can use the iterant algebra generated by $e$ and $\eta$ if we wish. \\

 We have the Dirac equation 
 $ \hat{E} = \alpha \hat{p} + \beta \hat{m}.$  
 Because the quantum operator for momentum is 
 $ \hat{p} = -i \partial/\partial x,$ the operator for energy is $\hat{E} = i\partial/\partial t,$ and the operator for mass is $\hat{m} = m,$ the Dirac equation becomes the
 differential equation below.
  $$i\partial \psi /\partial t = -i \alpha \partial \psi /\partial x + \beta m \psi.$$
  
 \noindent Let ${\cal O} = i\partial /\partial t + i \alpha \partial /\partial x - \beta m $ so that the Dirac equation 
 takes the form ${\cal O} \psi(x,t) = 0.$ \\
 
  \noindent  Now note that 
 ${\cal O} e^{i(px - Et)} = (E - \alpha p - \beta m) e^{i(px - Et)}.$\\
 
  \noindent  We let $\Delta = (E - \alpha p - \beta m)$ and let 
 $ U = \Delta \beta \alpha = (E - \alpha p - \beta m)\beta \alpha = \beta \alpha E + \beta p - \alpha m,$
 so that  $U^2 = -E^2 + p^2 + m^2 = 0.$\\
 
  \noindent   This nilpotent element leads to a (plane wave) solution to the 
 Dirac equation as follows:  We have shown that ${\cal O} \psi = \Delta \psi$ for 
 $\psi =  e^{i(px - Et)}.$  It then follows that 
 ${\cal O}(\beta \alpha \Delta \beta \alpha \psi) = \Delta \beta \alpha \Delta \beta \alpha  \psi =
 U^2 \psi = 0,$
  from which it follows that $\psi = \beta \alpha U e^{i(px - Et)}$ is a (plane wave) solution to the Dirac equation.
\bigbreak

This calculation suggests that we should multiply the operator ${\cal O}$ by $\beta \alpha$ on the right, obtaining the operator 
$${\cal D} = {\cal O} \beta \alpha =  i\beta \alpha \partial /\partial t + i \beta \partial /\partial x - \alpha m, $$
and the equivalent Dirac equation ${\cal D}\psi = 0.$ For the specific $\psi$ above we will now 
have ${\cal D} (U  e^{i(px - Et)} ) = U^2  e^{i(px - Et)} = 0.$ This idea of reconfiguring the Dirac equation
in relation to nilpotent algebra elements $U$ is due to Peter Rowlands \cite{Rowlands}.  Rowlands does this in the context of quaternion algebra.
Note that the solution to the Dirac equation that we have found is expressed in Clifford algebra. It can be articulated into specific vector solutions by using an iterant or matrix representation of the algebra.
\bigbreak

We see that $ U = \beta \alpha E + \beta p - \alpha m$ with $U^2 = 0$ is really the essence of  this plane wave solution to the Dirac equation. This means that a natural non-commutative algebra arises directly 
and can be regarded as the essential information in a Fermion. It is natural to compare this algebra structure with algebra of creation and annihilation operators that occur in quantum field theory.\\

\noindent If we let $\tilde{\psi} =  e^{i(px +Et)}$ (reversing time), then we have
$ {\cal D}\tilde{\psi} = (-\beta \alpha E + \beta p - \alpha m)\psi = U^{\dagger} \tilde{\psi},$
giving a definition of $U^{\dagger}$ corresponding to the anti-particle for $U\psi.$
\bigbreak

\noindent We have
$U =  \beta \alpha E + \beta p - \alpha m$
and
$U^{\dagger} =  - \beta \alpha E + \beta p - \alpha m.$\\

\noindent Note that here we have 
$(U + U^{\dagger})^2 =  (2 \beta p + \alpha m)^2 =  4 (p^2 + m^2 )= 4 E^2 ,$ and
$(U - U^{\dagger})^2 = - ( 2 \beta \alpha E)^2 = - 4 E^2 .$\\

\noindent We have that $U^{2} = (U^{\dagger})^{2} = 0 $ and $U U^{\dagger} + U^{\dagger} U = 4 E^{2}.$
Thus we have a direct appearance of the Fermion algebra corresponding to the Fermion plane wave solutions to the Dirac equation. Furthermore, the decomposition of $U$and $U^{\dagger}$ into the
corresponding Majorana Fermion operators corresponds to $E^2 = p^2 + m^2 .$\\

\noindent Normalizing by dividing by $2 E$ we have
$A =( \beta p + \alpha m)/E $ and 
$B = i \beta \alpha.$ so that
$A^2 = B^2 = 1$ and $AB + BA = 0.$ then
$U = (A + Bi)E$ and $U^{\dagger} = (A - Bi)E, $ showing how the Fermion operators are expressed in terms of the simpler Clifford algebra of Majorana operators (split quaternions once again).

\bigbreak

\subsection{Writing in the Full Dirac Algebra}
We have written the Dirac equation in one dimension of space and one dimension of time.
We now boost the formalism directly to three dimensions of space. We take an independent Clifford algebra generated by $\sigma_{1}, \sigma_{2}, \sigma_{3}$ with
$\sigma_{i}^{2} = 1$ for $i=1,2,3$ and $\sigma_{i}\sigma_{j} = - \sigma_{j}\sigma_{i}$ for 
$i \ne j.$ Now assume that $\alpha$ and $\beta$ as we have used them above generate an independent Clifford algebra that commutes with the algebra of the $\sigma_{i}.$ Replace
the scalar momentum $p$ by a $3$-vector momentum $p = (p_1 , p_2 , p_3 )$ and let 
$p \bullet \sigma = p_{1} \sigma_{1} +  p_{2} \sigma_{2} +  p_{3} \sigma_{3}.$ We replace
$\partial / \partial x$ with $\nabla  = (\partial /  \partial x_{1} , \partial / \partial x_{2}, \partial / \partial x_{2} )$
and $\partial p / \partial x$ with $\nabla \bullet p.$
\bigbreak

\noindent We then have the following form of the Dirac equation.
$i\partial \psi /\partial t = -i \alpha \nabla \bullet \sigma  \psi  + \beta m \psi.$\\

 \noindent Let ${\cal O} = i\partial /\partial t + i \alpha \nabla \bullet \sigma  - \beta m $ so that the Dirac equation 
 takes the form ${\cal O} \psi(x,t) = 0.$\\
 
 \noindent In analogy to our previous discussion we let 
 $\psi(x,t) =  e^{i(p \bullet x - Et)}$ and construct solutions by first applying the Dirac operator to this $\psi.$ The two Clifford algebras interact to generalize directly the nilpotent solutions and Fermion algebra that we have detailed for one spatial dimension to this three dimensional case. To this purpose the modified Dirac operator is
$$ {\cal D} = i\beta \alpha \partial/\partial t + \beta \nabla \bullet \sigma - \alpha m.$$
And we have that ${\cal D}\psi = U \psi$ where
$U = \beta \alpha E + \beta p \bullet \sigma - \alpha m.$
We have that $U^{2}= 0$ and  $U \psi$ is a solution to the modified Dirac Equation, just as before.
And just as before, we can articulate the structure of the Fermion operators and locate the corresponding Majorana Fermion operators.
 \bigbreak
 
 \subsection{Majorana Fermions}
There is more to do. We now discuss making  Dirac algebra distinct from the one generated by $\alpha, \beta, \sigma_1 , \sigma_2 , \sigma_3$ to obtain an equation that can have real solutions. This was the strategy that Majorana \cite{Majorana} followed to construct his Majorana Fermions. A real equation can have solutions that are invariant under complex conjugation and so can correspond to particles that are their own anti-particles. We will describe this Majorana algebra in terms of the split quaternions $\epsilon$ and $\eta.$ For convenience we use the matrix representation given below. The reader of this paper can substitute the corresponding iterants.

$$ \epsilon = \left(\begin{array}{cc}
			-1&0\\
			 0&1
			\end{array}\right),
			 \eta = \left(\begin{array}{cc}
			0&1\\
			1&0
			\end{array}\right).$$
Let $\hat{\epsilon}$ and $\hat{\eta}$ generate another, independent algebra of 
split quaternions, commuting with the first algebra generated by $\epsilon$ and $\eta.$
Then a totally real Majorana Dirac equation can be written as follows:
$$(\partial/\partial t + \hat{\eta} \eta \partial/\partial x + \epsilon \partial/\partial y + \hat{\epsilon} \eta \partial/\partial z - \hat{\epsilon} \hat{\eta} \eta m) \psi = 0.$$
\bigbreak

\noindent To see that this is a correct  Dirac equation, note that
$$\hat{E} = \alpha_{x} \hat{p_{x}} +  \alpha_{y} \hat{p_{y}} +  \alpha_{z} \hat{p_{z}} + \beta m$$
(Here the ``hats'' denote the quantum differential operators corresponding to the energy and momentum.)
will satisfy $$\hat{E}^{2} =  \hat{p_{x}}^{2} + \hat{p_{y}}^{2} + \hat{p_{z}}^{2} + m^{2}$$ if the algebra
generated by $\alpha_{x}, \alpha_{y}, \alpha_{z}, \beta$ has each generator of square one and each distinct pair of generators anti-commuting. From there we obtain the general Dirac equation by replacing
$\hat{E}$ by $i\partial/\partial t$, and $\hat{p_{x}}$ with $-i\partial/\partial x$ (and same for $y,z$).
$$(i\partial/\partial t +i\alpha_{x}\partial/\partial x +i\alpha_{y} \partial/\partial y +i\alpha_{z} \partial/\partial y - \beta m) \psi = 0.$$ 
This is equivalent to
$$(\partial/\partial t  +\alpha_{x}\partial/\partial x  + \alpha_{y} \partial/\partial y + \alpha_{z} \partial/\partial y +i \beta m) \psi = 0.$$
Thus, here we take $$\alpha_{x} = \hat{\eta} \eta, \alpha_{y} =  \epsilon, \alpha_{z} = \hat{\epsilon} \eta ,
\beta = i\hat{\epsilon} \hat{\eta} \eta,$$ and observe that these elements satisfy the requirements for the Dirac algebra. Note how we have a significant interaction between the commuting square root of minus one ($i$) and the element $\hat{\epsilon} \hat{\eta}$ of square minus one in the split quaternions. This brings us back to our original considerations about the source of the square root of minus one. Both viewpoints combine in the element $\beta = i \hat{\epsilon} \hat{\eta} \eta$ that makes this Majorana algebra work. Since the algebra appearing in the Majorana Dirac operator is constructed entirely from two commuting copies of the split quaternions, there is no appearance of the complex numbers, and when written out in $2 \times 2$ matrices we obtain coupled real differential equations to be solved. This is a beginning of a new study of Majorana Fermions. For more information about this viewpoint, see \cite{KR}. In the next section we rewrite the Majorana Dirac operator, guided by nilpotents, obtaining solutions that directly use the Majorana Fermion operators.\\		
		
\section{Nilpotents, Majorana Fermions and the Majorana-Dirac Equation}

Let ${\cal D} = (\partial/\partial t + \hat{\eta} \eta \partial/\partial x + \epsilon \partial/\partial y + \hat{\epsilon} \eta \partial/\partial z - \hat{\epsilon} \hat{\eta} \eta m). $
In the last section we have shown how  ${\cal D}$ can be taken as the Majorana operator through which we can look for real solutions to the Dirac equation.
Letting $\psi(x,t) =  e^{i(p \bullet r - Et)},$  we have
$${\cal D}\psi =( -iE + i(\hat{\eta} \eta p_{x} + \epsilon p_{y} + {\hat \epsilon} \eta p_{z}) - {\hat \epsilon}{\hat \eta} \eta m)\psi.$$

\noindent Let $$\Gamma = ( -iE + i(\hat{\eta} \eta p_{x} + \epsilon p_{y} + {\hat \epsilon} \eta p_{z}) - {\hat \epsilon}{\hat \eta} \eta m)$$ and 
$$U = \epsilon \eta \Gamma =  ( i(- \eta \epsilon E -\hat{\eta} \epsilon p_{x} + \eta p_{y} - \epsilon {\hat \epsilon} p_{z}) + \epsilon {\hat \epsilon} {\hat \eta} m).$$

\noindent The element $U$ is nilpotent, $U^2 = 0,$ and we have that
$U = A + iB,$
$AB + BA = 0,$
$A =  \epsilon {\hat \epsilon} {\hat \eta} m,$
$B=  - \eta \epsilon E -\hat{\eta} \epsilon p_{x} + \eta p_{y} - \epsilon {\hat \epsilon} p_{z},$
$A^2 = -m^2,$ and 
$B^2 = -E^2 + p_{x}^2 + p_{y}^2 + p_{z}^2 = -m^2.$\\

\noindent Letting $\nabla = \epsilon \eta {\cal D},$ we have a new Majorana Dirac operator with $\nabla \psi = U \psi$ so that
$\nabla (U\psi) = U^2 \psi = 0.$
Letting $\theta = (p \bullet r - Et),$ we have 
$U \psi = (A + Bi)e^{i \theta} = (A + Bi)(Cos(\theta) + i Sin(\theta)) =$
$(A Cos(\gamma) -BSin(\theta)) + i(B Cos(\theta) + A Sin(\theta)).$\\

\noindent Thus we have found two real solutions to the Majorana Dirac Equation:
$\Phi = A Cos(\theta) - B Sin(\theta), $
$\Psi = B Cos(\theta) + A Sin(\theta)$ with
$\theta= (p \bullet r - Et)$  and $A$ and $B$ the Majorana operators
$A =  \epsilon {\hat \epsilon} {\hat \eta} m,$
$B=  - \eta \epsilon E -\hat{\eta} \epsilon p_{x} + \eta p_{y} - \epsilon {\hat \epsilon} p_{z}.$\\

Note how the Majorana Fermion algebra generated by $A$ and $B$ comes into play in the construction of these solutions.
This answers a natural question about the Majorana Fermion operators. 
Should one take the Majorana operators themselves seriously as representing physical states? Our calculation suggests that one should take them seriously.  \\

In other work \cite{MajoranaBraid,MLogic,Iterants,Kauff:KP} we review the main features of recent applications of the Majorana algebra and its relationships with representations of the braid group and with topological quantum computing. The present analysis of the Majorana Dirac equation first appears in our paper \cite{KR}.\\		

\subsection{Spacetime in $1 + 1$ dimensions.}
Using the method of this section and spacetime with one dimension of space ($x$), we can write a real Majorana Dirac operator in the form
$\partial/\partial t + \epsilon \partial/\partial x + \epsilon \eta m$ where, the matrix representation is now two dimensional with 
$$ \epsilon = \left(\begin{array}{cc}
			-1&0\\
			 0&1
			\end{array}\right),
			 \eta = \left(\begin{array}{cc}
			0&1\\
			1&0
			\end{array}\right),
			\epsilon \eta  = \left(\begin{array}{cc}
			0&-1\\
			1&0
			\end{array}\right).$$
We obtain a nilpotent operator, ${\cal D}$ by multiplying by $i \eta:$
${\cal D} = i \eta \partial/ \partial t + i \eta \epsilon \partial / \partial x - i \epsilon m.$\\

\noindent Letting $\psi = e^{i(px - Et)},$ we have ${\cal D} \psi = (A + i B)\psi$ where
$A = \eta E + \epsilon \eta p$ and 
$B = - \epsilon m.$\\

Note that $A^2 = E^2 - p^2 = m^2$ and $B^2 = m^2,$ from which it is easy to see that $A + i B$ is nilpotent. $A$ and $B$ are the Majorana operators for this decomposition.
Multiplying out, we find 
$$(A + iB) \psi = (A + i B)(Cos(\theta) + i Sin( \theta)) = $$
$$(A Cos(\theta) - B Sin (\theta)) + i( B Cos(\theta) + A Sin (\theta))$$
where $\theta = px - Et.$
We now examine the real part of this expression, as it will be a real solution to the Dirac equation. The real part is
$$A Cos(\theta) - B Sin (\theta) = (\eta E + \epsilon \eta p) Cos(\theta) + em Sin(\theta) 
= \left(\begin{array}{cc}
			-m Sin(\theta) & (E-p) Cos(\theta)\\
			(E+p) Cos(\theta) & m Sin(\theta)
			\end{array}\right).$$
Each column vector is a solution to the original Dirac equation corresponding to the operator $$\nabla = \partial/\partial t + \epsilon \partial/\partial x + \epsilon \eta m$$
written as a $2 \times 2$ matrix differential operator. We can see this in an elegant way by changing to light-cone coordinates:
$ r = \frac{1}{2}(t + x), l =  \frac{1}{2}(t - x).$ (Recall that we take the speed of light to be equal to $1$ in this discussion.) Then
$\theta = px - Et = -(E-p)r - (E + p)l.$\\

\noindent The Dirac equation
$(\partial/\partial t + \epsilon \partial/\partial x + \epsilon \eta m)  \left(\begin{array}{c}
			\psi_{1}\\
			\psi_{2}
			\end{array}\right) = 0$ becomes the pair of equations
$$\partial \psi_{1}/ \partial l = m \psi_{2},$$
$$\partial \psi_{2}/ \partial r= - m \psi_{1}.$$
Note that these equations are satisfied by 
$$\psi_{1} = - m Sin(-(E-p)r - (E + p)l),$$ 
$$\psi{_2} = (E+p)Cos(-(E-p)r - (E + p)l)$$
exactly when  $E^2 = p^2 + m^2$ as we have assumed.
It is quite interesting to see these direct solutions to the Dirac equation emerge in this $1 + 1$ case. The solutions are fundamental and they are distinct from the 
usual solutions that emerge from the Feynman Checkerboard Model \cite{Feynman,KN:Dirac}. It is the above equations that form the basis for the Feynman Checkerboard model that is obtained by examining paths in a discrete
Minkowski plane generating a path integral for the Dirac equation.\\

\noindent {\bf Remark.} Note that a simplest instance of the above form of solution is obtained by writing 
$$e^{i(r + l)} = cos(r+l) + i sin(r+l) = \sum_{n=0}^{\infty} (\sqrt{-1})^{n} \sum_{i + j = n} \frac{r^{i}}{i!}\frac{l^{j}}{j!}.$$ 
Then with $\psi_{2} = cos(r+l)$ and $\psi_{1} = sin(r+l)$ we have
$\partial \psi_{1}/ \partial l =  \psi_{2},$
$\partial \psi_{2}/ \partial r= - \psi_{1},$
solving the Dirac equation in the case where $m=1.$\\

\noindent {\bf Remark.}  
Let $\psi_{R} = \sum_{k=0}^{\infty} (-1)^{k} \frac{r^{k+1}}{(k+1)!} \frac{l^{k}}{k!},$
$\psi_{L} = \sum_{k=0}^{\infty} (-1)^{k} \frac{r^{k}}{k!} \frac{l^{k+1}}{(k+1)!},$
$\psi_{0} = \sum_{k=0}^{\infty} (-1)^{k} \frac{r^{k}}{k!} \frac{l^{k}}{k!}.$
Then $\psi_{1} = \psi_{0}+\psi_{L}$ and $\psi_{2} = \psi_{0}-\psi_{R}$
give a solution to the Dirac equation in light cone coordinates as we have written it above with $m=1:$
$\partial \psi_{1}/ \partial l =  \psi_{2}, \partial \psi_{2}/ \partial r= - \psi_{1}.$
These series are shown in \cite{KN:Dirac} to be a natural limit of evaluations of sums of discrete paths on the Feynman Checkerboard. 
The key to our earlier approach is that if one defines
$$C[\Delta]^{x}_{k} = \frac{(x)(x-\Delta)(x-2 \Delta)\cdots(x-(k-1)\Delta)}{k!},$$
Then $C[\Delta]^{x}_{k}$ takes the role of $\frac{x^{k}}{k!}$ for discrete different derivatives with step length $\Delta$ and it can be interpreted as a 
choice coefficient. A Feynman path on a rectangle in Minkowski space can be interpreted as two choice of $k$ or $k+1$ points along the $r$ and $l$ edges of the 
rectangle. Thus the products in the limit expressions of the form $\frac{r^{k}}{k!} \frac{l^{k+1}}{(k+1)!}$ or $\frac{r^{k}}{k!} \frac{l^{k}}{k!}$ correspond to paths on the Checkerboard with 
$k$ corners in a limit where there are infinitely many such paths. The details are in our paper \cite{KN:Dirac}.
The solutions we have given above, motivated by the Majorana algebra are related in form to these path sum solutions.
We will investigate the relationship of this approach with the Checkerboard model in
a separate paper.\\

\end{document}